\documentclass[12pt]{iopart}

\usepackage{graphicx}
\usepackage{epsfig}
\usepackage{iopams}  
\usepackage{bm}
\usepackage{bbm}
\usepackage{graphics}
\usepackage{cite,hyperref}
\usepackage{color}

\begin{document}
\title{A Lie algebraic approach to a nonstationary atom-cavity system}
\author{C. A. Gonz\'alez-Guti\'errez}
\address{Instituto de Ciencias F\'isicas, Universidad Nacional Aut\'onoma de M\'exico, Apdo. Postal 48-3, Cuernavaca, Morelos 62251, M\'exico}
\ead{carlosag@fis.unam.mx, carlosgg04@gmail.com}
\author{O. de los Santos-S\'anchez}
\address{Instituto de Ciencias F\'isicas, Universidad Nacional Aut\'onoma de M\'exico, Apdo. Postal 48-3, Cuernavaca, Morelos 62251, M\'exico}
\ead{octavio.desantos@gmail.com}
\author{R. Rom\'an-Ancheyta}
\address{Instituto de Ciencias F\'isicas, Universidad Nacional Aut\'onoma de M\'exico, Apdo. Postal 48-3, Cuernavaca, Morelos 62251, M\'exico}
\ead{ancheyta6@gmail.com}
\author{M. Berrondo}
\address{Department of Physics and Astronomy, Brigham Young University, Provo UT 84602, USA}
\ead{berrondo@byu.edu}
\author{J. R\'ecamier}
\address{Instituto de Ciencias F\'isicas, Universidad Nacional Aut\'onoma de M\'exico, Apdo. Postal 48-3, Cuernavaca, Morelos 62251, M\'exico}
\ead{pepe@fis.unam.mx}

\begin{abstract}
In this work we study the generation of photons inside an ideal cavity with resonantly oscillating boundaries in the presence of a two-level atom. We make use of Lie algebraic techniques to obtain an approximate time-evolution operator and evaluate not only the resonant and dispersive regimes but also explore different regions of parameters. We have found a very good agreement between our approximate results and those obtained by numerical means. 
\end{abstract}
\maketitle

\section{Introduction}
 Creation of photons from vacuum fluctuations is one of the many fascinating effects in quantum theory. 
In a seminal paper dealing with the quantum theory of linearly polarized light propagating in a one dimensional cavity bounded by two ideal, infinite, parallel, plane mirrors which move with arbitrary trajectories, Moore \cite{moore} predicted the creation of real photons generated from vacuum due to  nonadiabatic variations in the boundary conditions of the field. 
This effect is now known as the dynamical Casimir effect (DCE); See Refs.~\cite{review-dodonov,rmodf} for recent reviews on the status of the DCE. For the generation of the effect,
it is necessary to rapidly modulate  the boundary conditions of the electromagnetic field with velocities close to the speed of light, which 
for a physical mirror,  may not be experimentally feasible.
In order to circumvent these difficulties, experiments with analogous  systems such as  superconducting circuits consisting of a coplanar transmission line resonators with tunable electrical length has been  performed. In these experiments the rate of change of the electrical length can be  done very fast by modulating the inductance of a superconducting quantum interference device (SQUID) at high frequencies \cite{expcasimir}. A different proposal, based on a trapped-ion implementation has been made recently  \cite{newjp}.

Interestingly, it was shown in \cite{dodonov-klimov} that one might expect a significant rate of photons generation inside ideal cavities with resonantly oscillating boundaries. The simplest model describing this effect takes into account a single resonant cavity mode whose frequency is rapidly modulated \cite{dodo2},  
  A quite different scenario can  take place when featuring a secondary system (a detector) inside a non-stationary cavity where the DCE is manifested. Along this line of research, the problem of the back action of different detectors on the rate of photon generation has been considered describing the detector as a two-level atom (or several atoms) \cite{dodonov,dodopra85} or by means of a harmonic oscillator tuned in resonance with the selected field mode \cite{pra8713,jpa4639}. More recent results have shown that DCE can  also manifest if one allow the Zeeman splitting of the qubit or the atom-field coupling to be time-dependent  \cite{liberato2009, dodonovjpa2014, sousa2015, veloso2015, dodonovpra2016}.

Here, we are in line with the aforementioned studies of tackling the problem of exploring the effect of adding a secondary system viewed as a two-level atom on the evolution of the field in a non-stationary cavity. To do this, we put forward an alternative approach  based on Lie-algebraic techniques since the constituent operators of the proposed unperturbed Hamiltonian model, the starting point of our treatment, turns out to generate a closed Lie algebra. This fact enables us to express the corresponding evolution operator of the whole system as product of exponentials according to the well-known Wei-Norman theorem, as will be outlined in section \ref{sec:1}. Our algebraic procedure, therefore, allows us to derive closed-form semi-analytical expressions for exploring, in section \ref{sec:numerics}, some quantities of physical interest to the problem at hand, such as the average value of Casimir photons, the variance of the field quadratures, and the evolution of the field on phase space in terms of its Q-function; the extent to which our approximate solutions are applicable is also stated by comparing them with the corresponding entirely numerical results. And finally, in section \ref{sec:conclusions}, some conclusions are given.
\section{The model and its approximate solution} \label{sec:1}

Consider the case of a closed cavity with a moving wall executing a periodic motion and a two-level atom inside it. The simplest Hamiltonian describing this system can be written as (in units of $\hbar$):

\begin{equation}\label{eq:H1}
\hat H  =  \omega(t)\hat n + \chi(t)\left(\hat a^2 +\hat a^{\dagger 2}\right)+\frac{\Omega}{2}\hat\sigma_z + g(\hat a +\hat a^{\dagger})(\sigma_{+} + \sigma_{-}), 
\end{equation}
where $\hat n$, $\hat a$, and $\hat a^{\dagger}$ are the usual number, annihilation, and creation operators, while $\hat{\sigma}_{z},\sigma_{\pm}$ refer to the Pauli matrices representing the two-level atom. We have chosen  $\chi(t) \equiv \frac{1}{4\omega(t)}\frac{d\omega(t)}{dt}$ and $\omega(t)=\omega_0\left[1+\epsilon \sin(\eta t)\right]$, with $|\epsilon| \ll \omega_0$ being the modulation amplitude and $\eta$ the frequency of the modulation. It is known that in the absence of the atom-field interaction, the mean number of photons grows exponentially if $\eta\simeq 2\omega_0$ and $\epsilon\omega_0 t\geq 1$ \cite{dodonov-klimov}.\\ 
  
Since the creation of photons is independent of the number operator, we take  $\omega(t)\simeq \omega_0$  and set the unperturbed Hamiltonian  as
\begin{equation}
\hat H_0 = \omega_0\hat n+\frac{\Omega}{2}\hat\sigma_z + \chi(t)\left(\hat a^2 +\hat a^{\dagger 2}\right).
\label{eq:H0}
\end{equation}
with the atom-field interaction Hamiltonian being given by
\begin{equation}
\hat V=g(\hat a +\hat a^{\dagger})(\hat\sigma_{+} + \hat\sigma_{-}).
\end{equation}
By virtue of the fact that the constituent field operators of $H_{0}$, namely, $\{ \hat{n}, \hat{a}^{\dagger 2}, \hat{a}^{2}\}$, form the basis of an su(1,1) Lie algebra, it is well known that the time evolution operator corresponding to such a Hamiltonian can be cast in terms of a product of  exponentials of the aforesaid operators \cite{wei}
\begin{equation}\label{eq:U0}
\hat U_0 = e^{-i\frac{\Omega}{2}t\hat\sigma_z}e^{\gamma_1 \hat n}e^{\gamma_2 \hat a^{\dagger 2}}e^{\gamma_3\hat a^2}e^{\gamma_4},
\end{equation}
with complex, time dependent functions $\gamma_i(t)$ to be determined  by solving the set of ordinary differential equations:
\begin{eqnarray}
\dot \gamma_1 = -i\omega_0 -4i e^{2\gamma_1}\chi(t)\gamma_2, \\
\dot \gamma_2 = \left(-i e^{-2\gamma_1}+4i e^{2\gamma_1}\gamma_2^2\right)\chi(t),\\
\dot \gamma_3=-i e^{2\gamma_1}\chi(t), \\
\dot \gamma_4 = -2i e^{2\gamma_1}\chi(t) \gamma_2.
\end{eqnarray}
The time evolution operator of the whole system in the interaction picture generated by $\hat U_0$  satisfies:
\begin{equation}
i\partial_t \hat U_I = \hat H_{I} \hat U_I,
\label{eq:Ui}
\end{equation}
with $\hat H_{I}=\hat U_0^{\dagger}\hat V\hat U_0 $ being the interaction picture Hamiltonian and $\hat U_I(t_0,t_0)=1$. This representation entails applying the Bogoliubov transformation to the annihilation and creation operators, namely, 
\begin{eqnarray*}
\hat U_0^{\dagger} \hat a \hat U_0 = t_1 \hat a + t_2 \hat a^{\dagger} , \\
\hat U_0^{\dagger} \hat a^{\dagger} \hat U_0 = t_3 \hat a + t_4 \hat a^{\dagger},
\end{eqnarray*}
where the transformation coefficients are such that $ t_1^{*} = t_4$ and $t_2^{*} = t_3 $ since the transformation is unitary, and are given in terms of the $\gamma_i$'s as follows:
\begin{eqnarray}
t_1 = e^{\gamma_1}-4e^{\gamma_1}\gamma_2\gamma_3, \\
t_2 = 2 e^{\gamma_1}\gamma_2, \\
t_3 = -2 e^{-\gamma_1}\gamma_3, \\
t_4 = e^{-\gamma_1}.
\end{eqnarray}
Thus, the interaction picture Hamiltonian takes the form:
\begin{equation}
\hat H_I  = g\left[((t_1 +t_3)\hat a +(t_2+t_4) \hat a^{\dagger})(\hat\sigma_{+}e^{i\Omega t} + \hat\sigma_{-}e^{-i\Omega t})\right].
\label{eq:Hi}
\end{equation}
At this point, let us keep only the terms that close a Lie algebra (an approximation that will be grounded on comparing its predictive accuracy with the corresponding numerical results based upon the whole Hamiltonian) to get a Jaynes-Cummings-type interaction of the form:
\begin{equation}
\tilde H_I \approx g\left[ \hat a\hat\sigma_{+}(t_1+t_3)e^{i\Omega t}+ \hat a^{\dagger}\hat\sigma_{-}(t_2+t_4)e^{-i\Omega t}\right],
\label{eq:Hiapprox}
\end{equation}
where operators that conserve the total number of excitations have been kept. So, we find it convenient to define the operators \cite{blas-moya,cordero,moya2,moyarep}
\begin{equation}
\hat b =\frac{\hat a \sigma_{+}}{\sqrt{M}}, \qquad  \hat b^{\dagger}=\frac{\hat a^{\dagger}\sigma_{-}}{\sqrt{M}},
\end{equation}
with $M = n+\frac{1}{2}(1+\sigma_z)$ being the total number of excitations in the corresponding ladder, so that the interaction Hamiltonian (\ref{eq:Hiapprox}) can be recast as:
\begin{equation}
\tilde H_I \approx \sqrt{M} g\left[ (t_1+t_3) \hat b e^{i\Omega t}+(t_2+t_4) \hat b^{\dagger} e^{-i\Omega t}\right].
\end{equation}

The number of excitations for a given ladder is a constant since the operators $\hat b$ and $\hat b^{\dagger}$ generate transitions between states of the same ladder (fixed $M= n+1$). The action of such operators upon the states $|e,n\rangle$ and $|g,n+1\rangle$ is given by:
\begin{equation*}
\hat b|e,n\rangle = 0, \quad \hat b|g,n+1\rangle = |e,n\rangle, \quad \hat b^{\dagger}|e,n\rangle = |g,n+1\rangle, \quad \hat b^{\dagger}|g,n+1\rangle = 0 ,
\end{equation*}
from which one can deduce the commutation relations:
\begin{equation*}
[\hat b, \hat b^{\dagger}] = \hat\sigma_z, \quad  [\hat\sigma_z,\hat b]=2\hat b, \quad  [\hat\sigma_z,\hat b^{\dagger}]=-2\hat b^{\dagger}.
\end{equation*}
Again, the Lie algebra generated by the set of operators $\{ \hat b, \hat b^{\dagger},\hat \sigma_z \}$ enables us to apply the Wei-Norman Theorem and to write the time evolution operator for the interaction part in a product form as:
 \begin{equation}\label{eq:UI}
 \hat U_I = e^{\beta_z \hat\sigma_z} e^{\beta_{+}\hat b^{\dagger}}e^{\beta_{-}\hat b}.
 \end{equation}
Substitution of this evolution operator into Schr\"odinger's equation (\ref{eq:Ui}) allows us to arrive at the following set of coupled differential equations for the complex, time-dependent functions $\beta_{z}(t)$ and $\beta_{\pm}(t)$:
\begin{eqnarray}
\dot \beta_z = -i e^{i\Omega t -2\beta_z} g\sqrt{M}(t_1+t_3) \beta_{+}, \\
\dot \beta_{+}= -i g\sqrt{M}\left(e^{-i\Omega t+2\beta_z}(t_2+t_4)+e^{i\Omega t-2\beta_z}(t_1+t_3) \beta_{+}^2\right), \\
\dot \beta_{-}=-i e^{i\Omega t -2\beta_z}g\sqrt{M} (t_1+t_3).
\end{eqnarray}
The solution to these equations must be obtained by numerical means. Nonetheless, having established the  evolution operator of the whole system, $\hat U=\hat U_{0}\hat U_{I}$, the algebraic scheme enables us to readily  proceed to the calculation of any physical observable. Let, for instance, the initial state of the system at time $t_{0}$ be $|\Psi(t_0)\rangle = \alpha |e,n\rangle+\beta|g,n+1\rangle$, where $|e\rangle$, $|g\rangle$ refer to the excited and ground atomic states and $|n\rangle$ corresponds to a field with $n$ photons.  Since the state is normalized we require $\vert\alpha\vert^2 +\vert\beta\vert^2=1$. At time $t$, the system has evolved into $|\Psi(t)\rangle =\hat U_0 \hat U_I |\Psi(t_0)\rangle$. Once obtained its explicit form, one can evaluate almost straightforwardly the average value of a given observable $\hat O(t)$: 
\begin{equation}
\langle \hat O(t)\rangle = \langle \Psi(t)|\hat O |\Psi(t)\rangle = \langle \Psi(t_0)|\hat U_I^{\dagger}\hat U_0^{\dagger}\hat O \hat U_0 \hat U_I|\Psi(t_0)\rangle. 
\label{eq:average}
\end{equation}

Consider, for example, the case where the operator $\hat O$ is the number operator. Application of $\hat U_I$ to the initial state $|\Psi(t_0)\rangle$ yields   
\begin{eqnarray} 
 |\Psi_I(t)\rangle& = &e^{\beta_z}(\alpha+\beta \beta_{-})|e,n\rangle + e^{-\beta_z}(\alpha \beta_{+}+\beta(1+\beta_{+}\beta_{-}))|g,n+1\rangle, \nonumber \\
& = & C_{e,n}|e,n\rangle + C_{g,n+1}|g,n+1\rangle,
 \end{eqnarray}
and the interaction picture representation of the number operator is 
\[ \hat n_I(t)= U_0^{\dagger}\hat n U_0 =(t_3 \hat a + t_4 \hat a^{\dagger})(t_1\hat a + t_2\hat a^{\dagger}), \]
which may be written as:
\begin{eqnarray}
\hat n_I(t)&=&(1-8\gamma_2\gamma_3)\hat n + 2\gamma_2 \hat a^{\dagger 2}+2\gamma_3(4\gamma_2\gamma_3-1)\hat a^2 -4\gamma_2\gamma_3 , \nonumber \\
& = & g_{11}\hat n +g_{20}\hat a^{\dagger 2}+g_{02}\hat a^2 +g_{00}.
\end{eqnarray}
So, based upon (\ref{eq:average}), the average value of the number operator is 
\begin{equation} 
\langle \Psi_I(t)|\hat n_I(t)|\Psi_I(t)\rangle = g_{11}\left[ n |C_{e,n}|^2 +(n+1)|C_{g,n+1}|^2\right]+g_{00},
\label{eq:npromg}
\end{equation}
where we have made use of the relation $|C_{e,n}|^2 +|C_{g,n+1}|^2 =1$ that follows from the normalization condition $\langle \Psi_{I}(t)| \Psi_{I}(t) \rangle =1$. We shall apply this prescription in the following section to explore the effectiveness of our semi-analytic approach, comparing our results with the entirely numerical solution for the whole Hamiltonian (\ref{eq:H1}).

\section{Semi-analytical and numerical results} \label{sec:numerics}

Let us now focus on reviewing to which extent the semi-analytical approach outlined above can be used to explore the effect of the atom on features of physical relevance concerning the dynamics of the cavity field, viz. the expectation value of the number of Casimir photons, the variances of its quadratures, and its evolution in phase space based upon the Q-function.\\

\begin{figure}[h!]
\begin{center}
\includegraphics[width=7.5cm, height=5cm]{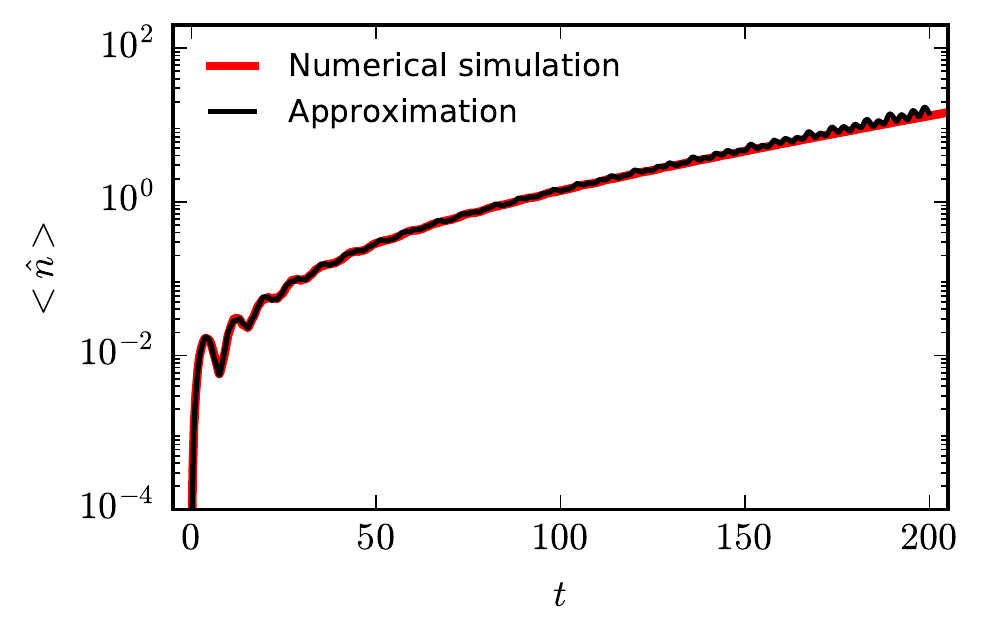} 
\includegraphics[width=7.5cm, height=5cm]{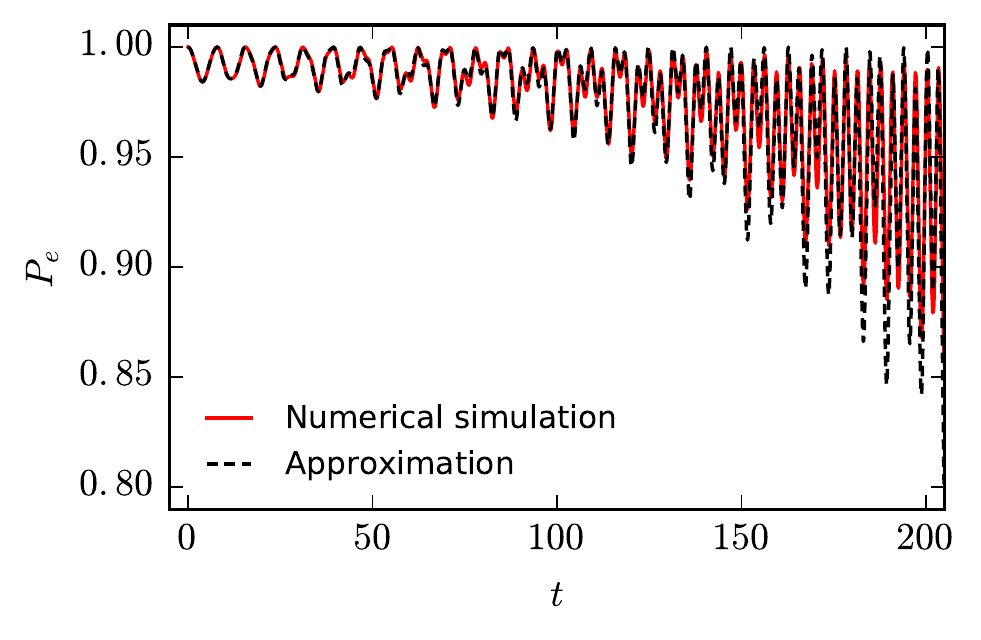} 
\caption{(Left panel) Average value of the number operator and (right panel) the probability of finding the atom in its excited state as functions of time. The system parameters are indicated in the inset (dispersive regime) and the initial state is taken to be $\rho_{AF}(0) = |e,0\rangle \langle e,0|$. Parameter set: $\{\Omega, \eta, \epsilon, g\}=\{0.2, 2.0, 0.02, 0.05\}\omega_{0}$.} 
\label{fig1}
\end{center}
\end{figure}

Since we are primarily interested in the aforesaid features when the state of the system at the initial time is $|\Psi(t_0)\rangle = \alpha |e,0 \rangle +\beta |g,1\rangle$, it is found that the average number of Casimir photons is given by
\begin{equation}
\langle \Psi_I(t)|\hat n_I(t)|\Psi_I(t)\rangle = g_{11} |\alpha \beta_{+}|^{2}e^{-2\Re \{ \beta_{z} \} }+g_{00},
\label{eq:npromp}
\end{equation}
 the influence of the atom-field interaction on photon creation is encapsulated in the time-dependent functions $\beta_{+}$ and $\beta_{z}$; needless to say, such an influence comes into play as long as there exists a nonzero probability of having the atom, initially, in its excited state ($\alpha \neq 0$). The outcome of Eq. (\ref{eq:npromp}) is depicted in the left panel of Fig. \ref{fig1} for the case $\alpha=1$, and the remaining parameters involved in the calculation are given in the inset and chosen in a way such that the atom-field interplay falls into the dispersive regime in which $\omega_{0} \gg \Omega$ and $|\Omega-\omega_{0}| \gg g$, say, $\Omega=0.2 \omega_{0}$ and $g=0.05$ (unless otherwise specified, we shall make use of red and black lines to label the numerical and approximate results, respectively, in subsequent descriptions). So, on the basis of the numerical result, we see that our approach works very well within the time interval displayed in the plot. Although a slightly oscillatory conduct is exhibited at the very beginning of the evolution, the archetypal exponential growth of such a quantity rules its overall profile. On the other hand, as far as the atom is concerned, the probability of finding it in its excited state turns out to be given by the expression 
\begin{equation} 
 P_e(t) = \frac{e^{2 \Re \{ \gamma_{4}+\beta_{z} \} }}{\sqrt{1-4|\gamma_{2}|^{2}e^{4\Re \{ \gamma_{1} \}}}},
 \end{equation}
which is calculated from the trace operation $\Tr \{ \hat{\rho}_{A} |e\rangle \langle e| \}$, with $\hat{\rho}_{A}=\Tr_{F} \{ \hat{\rho}_{AF}\}$ being the reduced density operator of the atom ($\Tr_{F}$ means tracing over the degree of freedom of the field) and $\rho_{AF}= |\Psi (t) \rangle \langle \Psi (t)|$ the density operator of the composite system. This quantity is displayed in the right panel of Fig. \ref{fig1}. One can see that its spiky behavior in the early stages of system's evolution is in agreement with the corresponding numerical result; this conduct is also in accord with the fact that the state of the atom, being out of resonance with the field, remains almost unchanged around the state it started up ($P_{e}(t)\approx 1$).

\begin{figure}[h!]
\begin{center}
\includegraphics[width=7.5cm, height=5cm]{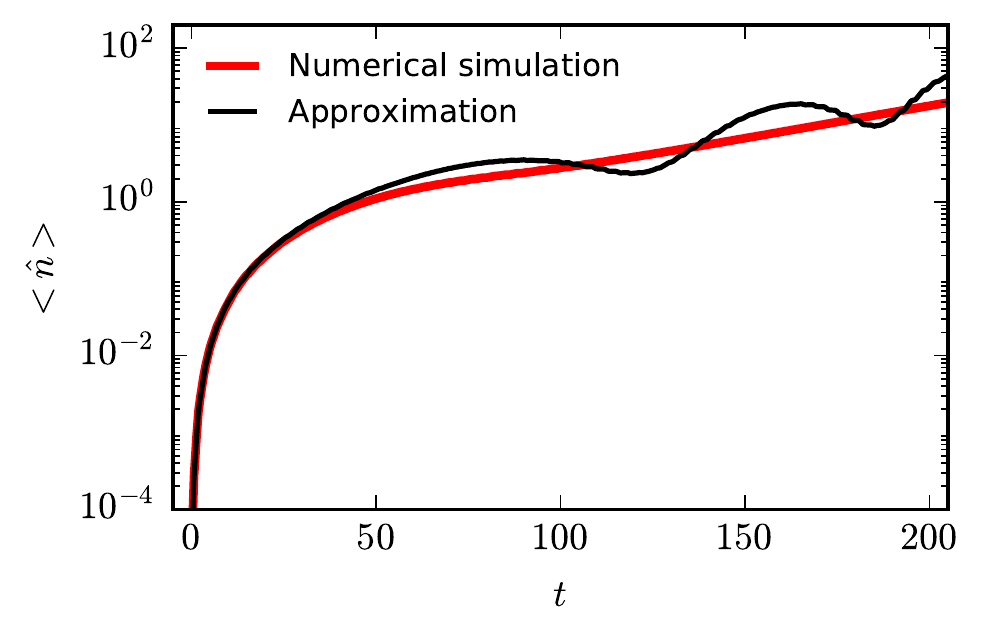} 
\caption{Average value of the number operator as a function of time. The system parameters are indicated in the inset (resonant regime) and the initial state is $|e,0\rangle$.
Parameter set: $\{\Omega, \eta, \epsilon, g\}=\{1.0, 2.0, 0.02, 0.02\}\omega_{0}$.} 
\label{fig2}
\end{center}
\end{figure}

While being a quite acceptable description in the dispersive regime, our semi-analytical approach starts loosing accuracy as the interaction between the atom and the field is in resonance ($\Omega=\omega_{0}$), as exemplified  in Fig. \ref{fig2} where the photon production is plotted as a function of time for the set of system parameters indicated in the inset. In this regime, one can see, again, a preponderant exponential growth in photon production, whereas our approximation starts displaying, as time elapses, a conspicuous oscillatory profile around the value given by the entirely numerical solution. \\

At this stage, it is worth mentioning that our semi-analytical approach goes beyond the algebraic treatment reported in Ref. \cite{dodonov} where the coupling strength of the cavity field with the atom is regarded as a small perturbation parameter in comparison with the modulation amplitude, i.e. $g \ll \epsilon$. Indeed, besides being applicable in this limit, the present approach encompasses both dispersive and resonant regimes and allows the aforesaid parameters to have values of the same order of magnitude (as those used for figure (\ref{fig2})).

\begin{figure}[h!]
\begin{center}
\includegraphics[width=7.5cm, height=5cm]{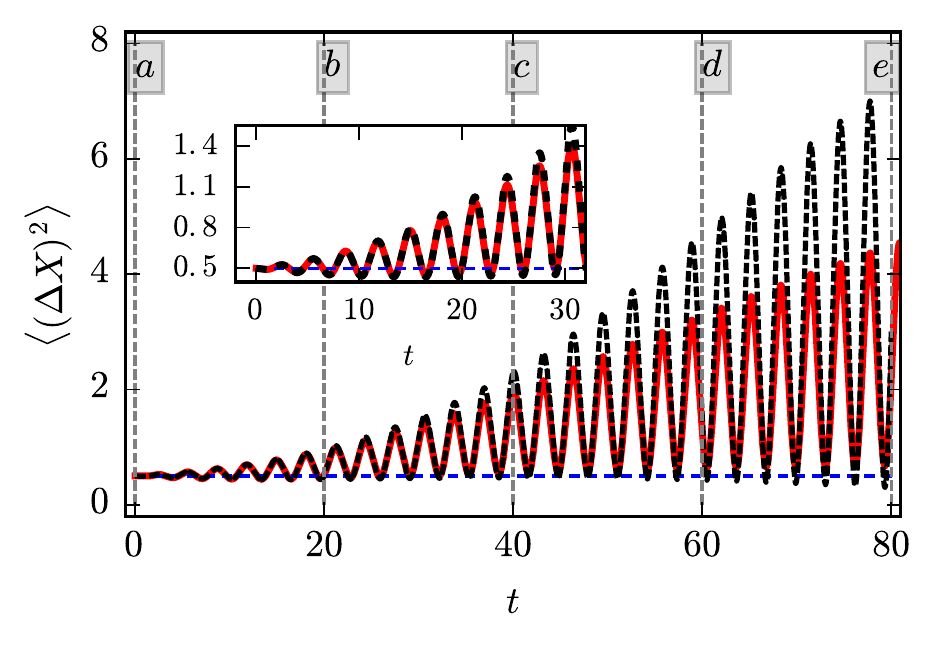} 
\includegraphics[width=7.5cm, height=5cm]{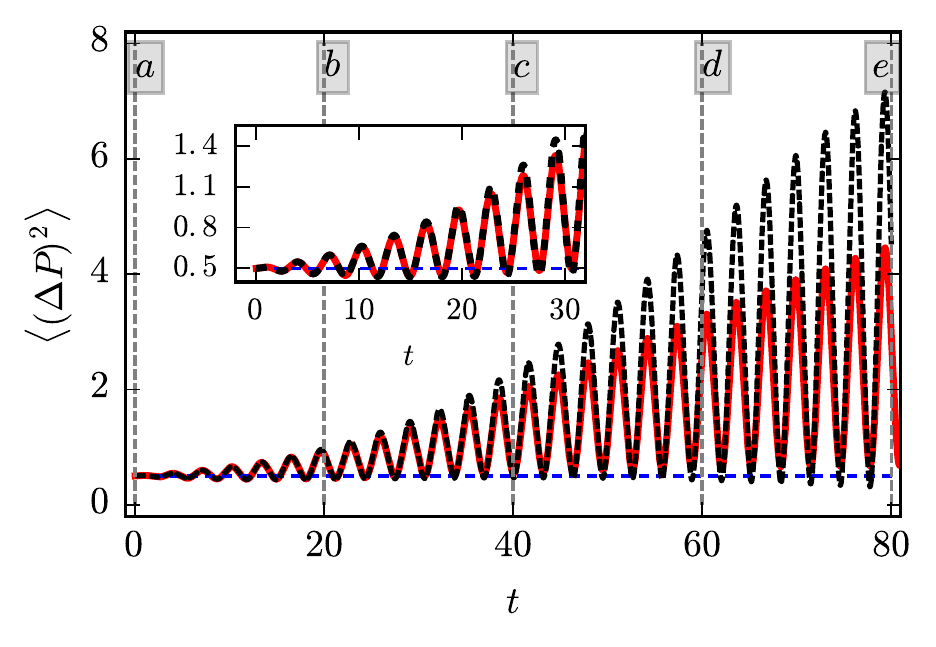} 
\caption{Plot of the variance of the field quadratures $\hat{X}=(\hat{a}+\hat{a}^{\dagger})/\sqrt{2}$ (left panel) and $\hat{P}=(\hat{a}-\hat{a}^{\dagger})/\sqrt{2}i$ (right panel) as functions of time for the same set of parameters as in Fig. \ref{fig2}. The inset focuses on the shortened time interval, $0 < t < 30$, within which a good agreement between the numerical (red line) and semi-analytical (black-dashed line) results can be observed.} 
\label{fig3}
\end{center}
\end{figure}

Before proceeding to explore the evolution of the field on phase space, let us compute the variances of the field quadratures, labeled as $\hat{X}=(\hat{a}+\hat{a}^{\dagger})/\sqrt{2}$ and $\hat{P}=(\hat{a}-\hat{a}^{\dagger})/\sqrt{2}i$, that can be determined from Eq. (\ref{eq:average}) by using the expressions $\langle \Psi _{I}(t) |\hat{a}^{\dagger}_{I} |\Psi_{I}(t)\rangle  =  (\beta \beta_{+}^{\ast}\alpha^{\ast}-2\beta^{\ast}\beta_{+}\alpha \gamma_{3} )e^{-2 \Re \{ \beta_{z} \}-\gamma_{1}}$ and $\langle \Psi _{I}(t) |\hat{a}^{\dagger 2}_{I} |\Psi_{I}(t)\rangle  =  - 2\gamma_{3}e^{-2\gamma_{1}} (2|\alpha \beta_{+}|^{2}e^{-2 \Re \{ \beta_{z} \} }+1 )$, together with (\ref{eq:npromp}). For the same set of parameters and initial conditions as in Fig. (\ref{fig2}), the outcome of the variance $\langle (\Delta \hat{X})^{2} \rangle = \langle \hat{X}^{2} \rangle-\langle \hat{X} \rangle^{2}$ as a function of time is displayed in the left panel of Fig. \ref{fig3} along with the corresponding numerical result for the sake of comparison. Again, the short time behavior of the approximate result matches that of the converged numerical one as seen in the inset of the figure, reveling a small degree of squeezing (i.e., $\langle (\Delta \hat{X})^{2} \rangle<1/2$, see the blue-dashed line in the figure as a reference) at certain values of time. Discrepancies between the approximate and the numerical results become noticeable for times larger than $t\simeq 40$, where, unlike the numerical outcome, our approximation still predicts squeezing in both quadratures around the interval $60 < t < 80$ and a more pronounced dispersion at certain periods of time. A similar behavior is observed for the dispersion in the conjugate quadrature $\hat{P}$ as seen in the right panel. \\

\begin{figure}[h!]
\begin{center}
\includegraphics{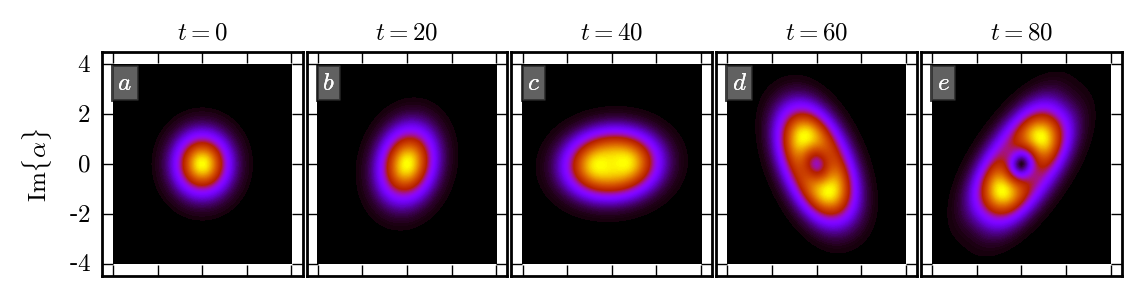} 
\includegraphics{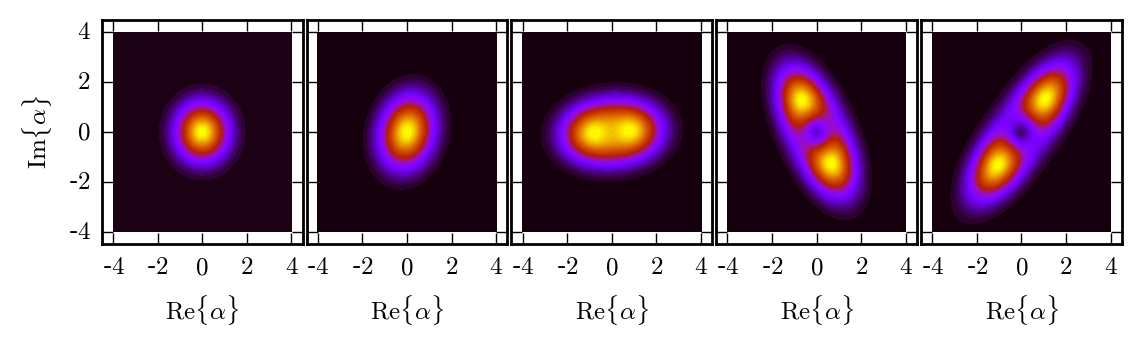} 
\caption{Evolution of the Q-function of the field density operator at time instants $t=0$, $20$, $40$, $60$, and $80$ when the initial state of the composite system is such that  $\rho_{AF}(0) = |e,0\rangle \langle e,0|$. The numerical and semi-analytical results correspond, respectively, to the upper and lower rows. The set of system parameters is the same as in Fig. \ref{fig2} (resonant regime).} 
\label{fig4}
\end{center}
\end{figure}

In order to  make a link between the previous results and the evolution of the field on phase space let us consider the Husimi Q-function, which is defined as the diagonal matrix element of the field density operator $\hat{\rho}_{F}= \Tr_{A} \{ \rho_{AF} \}$ (the trace is now over the atomic variables) between standard coherent states, i.e., 
\begin{equation}
Q(z) = \frac{1}{\pi}\langle z|\hat{\rho}_{F}|z\rangle,
\end{equation}
where $|z\rangle = \exp (-|z|^{2}/2) \sum_{k} z^{k}/\sqrt{k!} |k\rangle $. For the particular case we have been focused on ($\alpha=1$), substitution of $\hat{\rho}_{F}$ into this expression allows us to arrive at the sought result 
\begin{equation}
Q(z) = \frac{1}{\pi} e^{-|z|^{2}+2\Re \{ \gamma_{4}+z^{\ast 2} \gamma_{2}e^{2\gamma_{1}} \} } \left(e^{2 \Re \{ \beta_{z} \}}+|z\beta_{+}|^{2}e^{ 2\Re \{ \gamma_{1}-\beta_{z} \} } \right).
\end{equation}
This distribution function is portrayed as a density plot in the upper row of Fig. \ref{fig4} (the lower row corresponds to the numerical result) at the time instants $t=0$, $20$, $40$, $60$ and $80$, which, in turn, are indicated by the dashed-black vertical lines labeled as $a$, $b$, $c$, $d$ and $e$, respectively,  in Fig. \ref{fig3}. At $t=0$ the state of the field starts as a minimum-uncertainty state. It evolves and at say $t=20$, it attains a somewhat more elongated form along the vertical axis whose dispersion, given by $\langle (\Delta \hat{P})^{2} \rangle$, see right panel of Fig. \ref{fig3}, is duly quantified to be slightly above the coherent-state value. We also see that the presence of the atom foster the formation of what seems to be well localized two phase components (see frames (d) and (e)), which in fact begin to be barely  observable at $t=40$ (frame (c)). Although our approach assesses a wider spreading effect of the state's phase-space distribution, reflected in the dispersion relations of Fig. \ref{fig3}, than the one observed for the numerical outcomes for $t > 40$, it correctly predicts (qualitatively) the foregoing cat-like splitting behavior on short time scales. For a more detailed view of the field's evolution on phase space, see the supplemental material along with this work. \\

\begin{figure}[h!]
\begin{center}
\includegraphics[width=8cm, height=5cm]{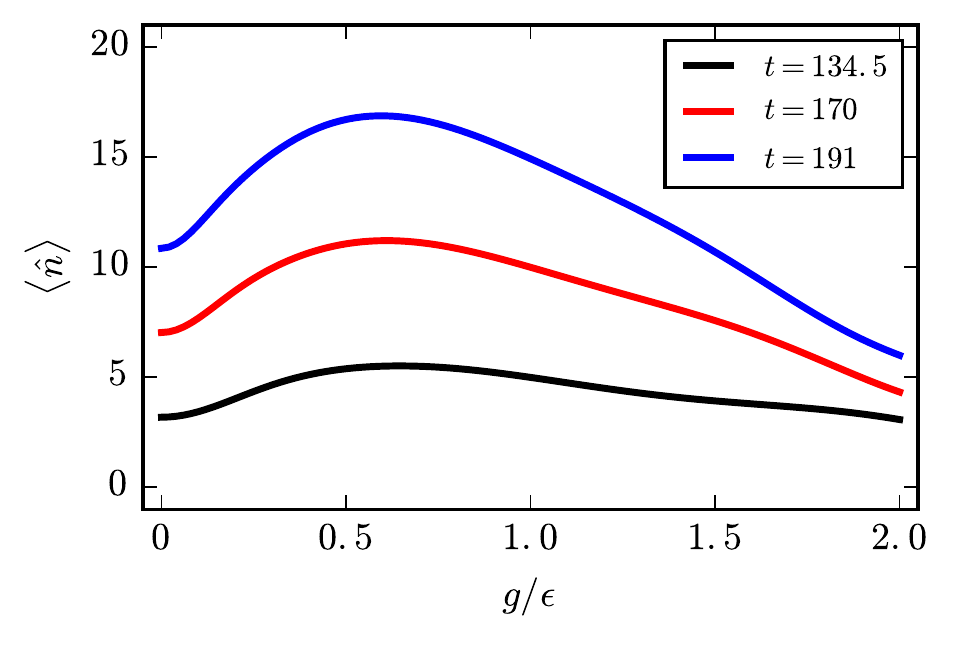} 
\caption{Average value of the number operator as a function of ratio $g/\epsilon$ evaluated at time instants $t=134.5$, $170$, and $191$, approximately. The initial state of the composite system is such that  $\rho_{AF} (0)= |e,0\rangle \langle e,0|$. Parameter set: $\{\Omega, \eta, \epsilon\}=\{1.0, 2.0, 0.02\}\omega_{0}$} 
\label{fig5}
\end{center}
\end{figure}

Finally, to close this section we find it pertinent to add a brief comment on the dependency of photon creation upon the coupling strength $g$ between the field and the atom that, up to our knowledge, had not yet been explored elsewhere. The overall behavior of the average photon number $\langle n \rangle$ is sketched in Fig. \ref{fig5}, in the resonant regime ($\Omega=\omega_{0}$, together with $\eta=2\omega_{0}$), at $t=134.5$, $170$, and $191$, as a function of the ratio $g/\epsilon$; the initial state of the whole system is $\rho_{AF}(0) = |e,0\rangle \langle e,0|$. These particular results were carried out by using a purely numerical procedure and revealed the importance of taking into consideration the existence of an interesting, albeit subtle, trade-off between the main system parameters involved in the process, $\epsilon$ and $g$, in the sense that particular values of their ratio can give rise to a discernible enhancement of Casimir photons (see, e.g., the profile at $t\approx 134.5$) which takes place as long as the counter-rotating terms are featured in the interaction  Hamiltonian (\ref{eq:Hi}).

\section{Concluding remarks and outlooks} \label{sec:conclusions}
In this work we have developed an approximate method for the study of a closed cavity with a moving wall executing periodic motion 
with a two-level atom inside the cavity. We take as unperturbed Hamiltonian that of the atom, the field and the part arising from the motion of the mirror. The interaction part of the Hamiltonian is that corresponding to the atom-field interaction which is treated within the rotating wave approximation and we thus have an interaction that preserves the total number of excitations. This approximation is valid whenever the atom-field coupling $g\ll \Omega$. In order to test the validity of our method we considered the same set of parameters as those used in Ref.~\cite{dodonov} where analytical expressions were obtained in the dispersive and the resonant limits, and we found a good agreement with their results. We also considered cases far from the resonance and the dispersive  regimes and we made a numerical evaluation in order to test our approximate results. We want to stress the fact that in the numerical evaluation we did not make use of the rotating wave approximation. One of the main effects due to the presence of the atom inside the cavity is an enhancement in the generation of Casimir photons, this is a function of the coupling parameter $g$ and it shows a maximum around $g/\epsilon\simeq 0.6$.

Finally, it is worth underlying the possibility of adapting the present algebraic scheme to carry out the assessment of quantum fluctuations of physical interest that, as far as we know, has not been undertaken in this context, such as time-dependent spectrum of light \cite{Eberly}, phase- and intensity-intensity correlation measurements \cite{book}, which involves the calculation of standard correlation functions, provided we restrict ourselves to the short-time behavior of the system. Another potential application would be to provide a protocol and/or scheme to the thermalization of the cavity through a random injection of atoms into it (see, e.g., Refs. \cite{Li,Liao,Quan,Ceren}) like the micromaser scenario; this would open up the possibility of implementing quantum heat engines in the context of non-stationary cavities.  A modest step toward tackling the problem of an empty non-stationary lossy cavity where the DCE in its bounded regime is manifested can be found in \cite{aoc}. 

\section{Acknowledgements} We acknowledge partial support from DGAPA-UNAM project PAPIIT IN113016, and we thank Reyes Garc\'{\i}a for the maintenance of our computers. CGG and RRA would like to express their gratitude to CONACyT-M\'exico for their respective scholarships. O de los SS wants to thank Professor J. R\'ecamier for his hospitality at ICF.

\end{document}